\documentclass[twocolumn,aps,prl,showpacs,superscriptaddress]{revtex4}

\usepackage{graphicx}
\usepackage{amsmath,amssymb}
\usepackage{color}

\newcommand{\ket} [1] {\vert #1 \rangle}

\newcommand{\be}{\begin{eqnarray}}
\newcommand{\ee}{\end{eqnarray}}

\newcommand{\beq}{\begin{equation}}
\newcommand{\eeq}{\end{equation}}
\newcommand{\beqa}{\begin{eqnarray}}
\newcommand{\eeqa}{\end{eqnarray}}
\newcommand{\sS}{\spadesuit}
\newcommand{\sH}{\textcolor[rgb]{1.00,0.00,0.00}{\heartsuit}}
\newcommand{\sD}{\textcolor[rgb]{1.00,0.00,0.00}{\diamondsuit}}
\newcommand{\sC}{\clubsuit}

\begin{document}


\title{Quantum bidding in Bridge}



\author{Sadiq Muhammad}
\affiliation{Department of Physics, Stockholm University, S-10691,
 Stockholm, Sweden}

\author{Armin Tavakoli}
 \affiliation{Department of Physics, Stockholm University, S-10691,
 Stockholm, Sweden}

\author{ Maciej Kurant}

\affiliation{Department of Information Technology and Electrical Engineering, ETH Zurich, Switzerland}

\author{Marcin Paw{\l}owski}

\affiliation{Instytut Fizyki Teoretycznej i Astrofizyki, Uniwersytet
Gda\'{n}ski, PL-80-952 Gda\'{n}sk, Poland}
\affiliation{Department of Mathematics, University of Bristol, Bristol BS8 1TW,
United Kingdom}

\author{Marek \.Zukowski}

\affiliation{Instytut Fizyki Teoretycznej i Astrofizyki, Uniwersytet
Gda\'{n}ski, PL-80-952 Gda\'{n}sk, Poland}

\author{Mohamed Bourennane}
 \affiliation{Department of Physics, Stockholm University, S-10691,
 Stockholm, Sweden}


\date{\today}
.


\begin{abstract}
Quantum methods allow to reduce communication complexity of some computational tasks, with several separated partners, beyond classical constraints.  Nevertheless, experimental demonstrations of this fact are thus far limited to some abstract problems, far away from real-life tasks.
We show here, and demonstrate experimentally, that the power of reduction of communication complexity can be harnessed to gain advantage in famous, immensely popular,  card game - Bridge.   The essence of a winning strategy in  Bridge is efficient communication between the partners. The rules of the game allow only  specific form of communication, of a very low complexity (effectively one has a strong limitations on number of exchanged bits).   Surprisingly, our quantum technique is not violating the existing rules of the game (as there is no increase in information flow). We show that our quantum Bridge auction corresponds to a biased nonlocal Clauser-Horne-Shimony-Holt (CHSH) game,  which is  equivalent to a $2\to 1$ quantum random access code. Thus our experiment is also a realization of such protocols. However, this correspondence is not full which enables the Bridge players to have efficient strategies regardless of the quality of their detectors.
\end{abstract}



\maketitle

\section{I. Introduction}
Quantum information science breaks the limitations of conventional information transfer, cryptography and computation. Communication complexity problems (CCPs)\cite{yao79} were  shown to have quantum protocols, which outperform any classical counterparts. In CCPs, two types can be distinguished. The first type minimizes the amount of information exchange necessary to solve a task with certainty \cite{cleve97,buhrman99,buhrman97}. The second type maximizes the probability of successfully solving a task with a restricted amount of communication \cite{buhrman97, hardy99, BZZ02}. Such studies aim, e.g., at a
speed-up of a distributed computation by increasing the communication efficiency, or at an optimization of VLSI circuits and data structures \cite{kushilevitz97}.

Fundamentally, there exist connections between quantum CCPs, quantum games, and tests of the foundations of quantum mechanics. It has been shown that for every CCP there is a corresponding quantum game and vice versa. Furthermore, it has also been proved that for every Bell inequality and for a broad class of protocols, there always exists a multi-party CCP, for which the protocol assisted by quantum states, which
violate the Bell inequality, is more efficient than any classical protocol \cite{BZZ02,BZPZ04}. However, in contrast with cryptography, the existing demonstrations of quantum protocols reducing communication complexity are abstract problems, and have no practical applications. Our aim is to apply quantum correlations in protocols related to a well known real-life task. The task that we are considering here is playing the card  game of (Duplicate) Bridge.

The essence of a successful game of Bridge is an efficient communication between the partners. Due to the rules of the game, the form and the amount of information exchanged between the partners is severely restricted. We show that using the quantum resources the players can increase their winning probability. What is important, our protocol does not require any change in the rules of the game (or perhaps the very existence of it would force a change in the rules, so as to add a missing point: no quantum reduction of communication complexity is allowed during the game). In order to use our scheme, the players need to share an entangled state and locally measure its subsystems. Such procedure is not against the rules of the World Bridge Federation \cite{Bridgerules}, and it is not a method of transferring information (no `Bell-telephone' rule).  Our aim is to show, that one can exploit the difference between the quantum and the classical resources in CCPs, in which there are strict limits on the amount of communication, to gain advantage in Bridge. So, whenever the rules governing a real-life situation put a limit on the amount of communication, it should also be specified whether quantum resources are allowed or not, since the quantum protocols, which outperform the classical counterparts, are within the reach of the current state-of-the-art technology, as shown by our experiment. We present an experimental realization of a quantum Bridge protocol, in which the quantum resources provide an advantage over the classical.
\section{II. Bridge}
Bridge is one of the world's most popular card games. It is a trick-taking game with  the  standard deck of 52 cards. The game is  played by four players playing in pairs with partners sitting at opposite sides of  a table, and named as West ($W$)-East ($E$) and
North ($N$)-South ($S$). The game consists of several deals each progressing through two phases: the auction (called bidding),
and playing the hand (called trick-taking) . The bidding phase starts with the dealer and rotates around the table clockwise with each player making a bid for the contract. The team that wins the auction undertakes to win a certain number of tricks during the game.
The second phase of the game is the standard trick-taking play.  Here, the winners of the auction try to fulfill their contract (taking as
many tricks as the contract obliges them to) while the opposing team intends to prevent them from doing so. The team that succeeds with its
task wins the round.
\begin{figure}[t]
\centerline{\includegraphics[width=0.82\columnwidth]{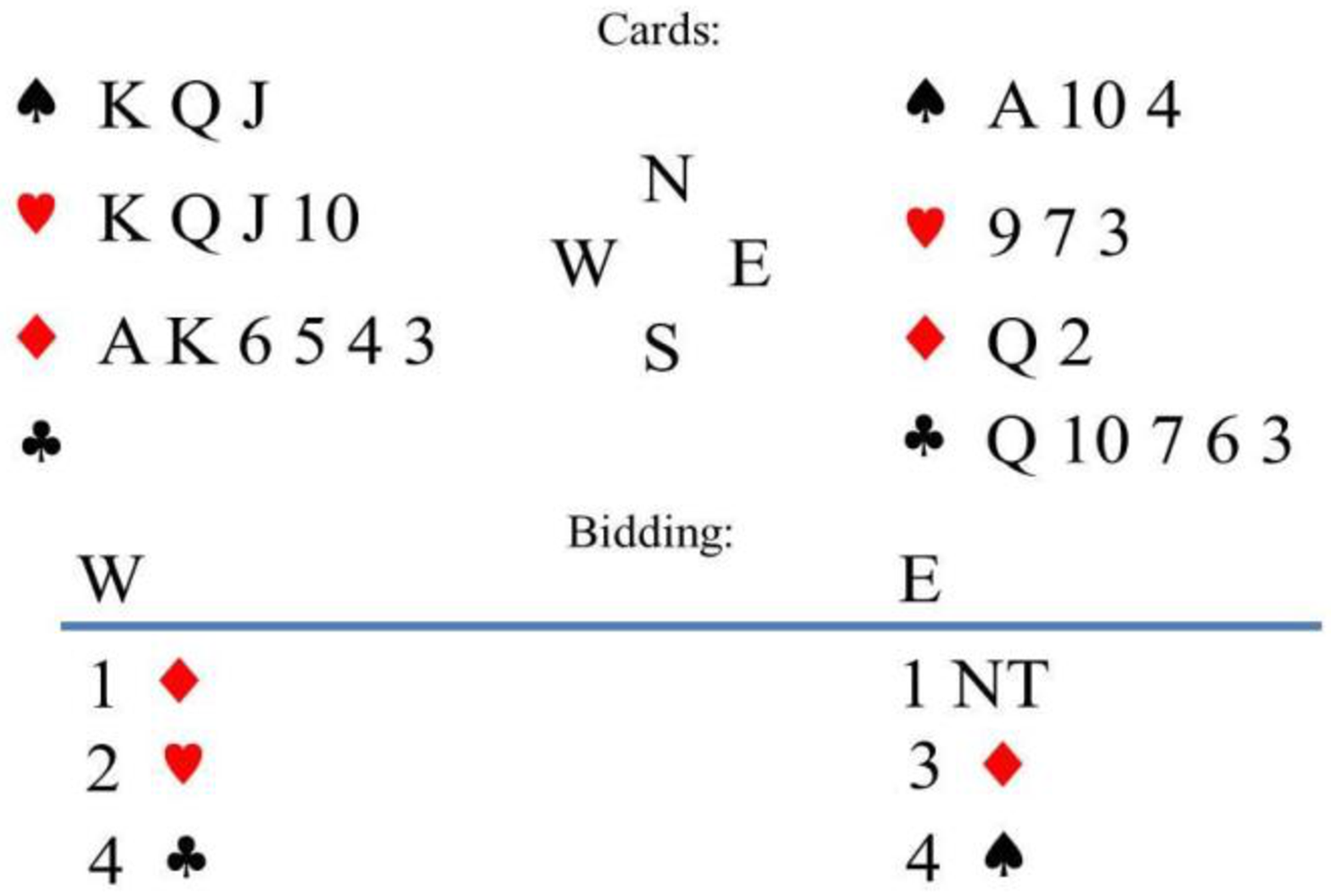}}
\caption{Cards and Bidding for Bridge. An example of cards and bidding for the $W$ and $E$ partners.}
\label{cards-Bridge}
\end{figure}
In both phases of Bridge, efficient communication is of vital importance for the partnership. This is even more true in the tournament version of the game -  duplicate Bridge recognized by the International Olympic Committee as a sport. There, a larger group of partnerships play against each other with prearranged hands to reduce the factor of chance to a minimum. What counts in this variation is earning more points with a given hand than the other partnerships had. This is done by winning the optimal contract and flawless play. To find the optimal contract both partners need to exchange information about their hands. This is done during the auction by bidding. Each bid gives the other player some information but, as in all auctions, the next bid has to be higher. This means that the more information is exchanged the higher the contract will be and more difficult to make. In order to optimize the communication between the partners during the auction phase various communication protocols have been devised. A short description of duplicate bridge is given in the appendix. More can be found in any of the numerous books on this subject, e.g.\cite{Bridge}.

In this paper we aim to show that there are quantum communication strategies outperform the classical ones in the game of bridge. Unfortunately, there is no best classical strategy. We have contacted Tommy Gullberg a World Life Master in duplicate bridge, author of many articles and books about the game, who told us: “A professional player plays with a strategy of his own preference since he finds it optimal. Thus if you go to a professional tournament you will encounter a wide range of different strategies, all with some strengths and some weaknesses, since there clearly is no such thing as an optimal classical bridge strategy"\cite{tommy}.

However, all the strategies must involve a "slam-seeking convention", which is a sub-strategy that players use if they have very strong cards hope to win a big bonus. Although, quantum strategies are probably useful at many different situations in bridge, in this paper we only give an example for improving Roman Key-card Blackwood (RKB)\cite{blackwood}. It is a very popular slam-seeking strategy and often used in bridge games today. If you play a game of bridge (in a tournament) it is very likely that the RKB will be used, probably several times. Therefore, the optimal classical strategy is likely to include it\cite{tommy}. And even if it does not, other slam-seeking strategies like Gerber \cite{gerber} differ from it only slightly and the quantum protocol can be modified accordingly. To show that the problem at which the quantum protocols excel at is ubiquitous in bridge in the appendix we present how the strategy used for RKB can be modified to be used in playing the hand.

The following paragraph plunges deep into the nuances of bidding. A reader not interested in the game should skip it and know only this: The task for the players is to decide if they have cards strong enough to try to get a bonus. To this end, one of them will ask questions (encoded in bids) and the other provide answers. In this particular case there are two important questions but only one of them can be asked. Classical and quantum ways of dealing with this problem are described in the next section.

Consider the cards and bidding in a bridge scenario shown in figure  \ref{cards-Bridge}. The bidding can be explained as follows, W looks at his cards. He has good cards in diamonds ($\sD$) and therefore makes the bid $1\sD$, suggesting that the partnership can win $1$ more trick than the minimum (which is always defined as $6$ out of $13$ possible). E does not have particularly good cards in any suit and therefore gives the answering bid of $1NT$ i.e., he suggests $7$ tricks without trump suit. By this bid, W understands that E has no significant strength in $\sD$ and does not have more than $4$ cards in any suit (otherwise he would have suggested that suit). Therefore W suggests his second best suit, $\sH$. E now knows that W prefers $\sD$ but can play with $\sH$ if necessary. E has more strength in $\sD$ and hence suggests $3\sD$ i.e, $9$ tricks with $\sD$ suit as trumps. The partnership has now settled the trump suit. Now they undertake a more difficult contract (by making more bids) in exchange for more information about the partner's hand.  W makes a so called cuebid of $4\sC$, directly suggest $10$ tricks but the cuebid implies that W is aiming for $12$ tricks. By choosing $\sC$, W indicates that he has either no cards at all or one strong card in the suit of $\sC$. W is thus asking E with his bid if E has any key cards (defined as four aces and trump King). E's answer says that he indeed has the key cards needed. However W has no information about the last missing card of interest, the queen of trumps $\sD$Q,
 Unfortunately, we have no standard tools to ask about $\sD$Q without exceeding the safety level of 5$\sD$. A typical Bridge strategy is Roman Key-card Blackwood 4NT (key-card refers to the king of trumps and the four aces; NT stands for no trump)\cite{blackwood}. If one partner wishes to ask the other whether he/she has the queen of trumps in his hand he calls out a bid on which meaning the players have agreed upon.
The standard answers for the partner are: For a bid $5\clubsuit$ means that the partner has $0$ or $3$ key cards; similarly $5\sD$, $5\sH$, and $5\spadesuit$ means $1$ or $4$ key cards, $2$ or $5$ key cards without the trump queen, and $2$ or $5$ key cards with the trump queen respectively.  Clearly, there is no way to check for the trump queen without exceeding the level of $5\sD$.Thus the classical bidding techniques are useless, and the partner $W$  may be forced to guess whether the partner $E$ has the trump queen or not.

\section{III. Bridge as communication complexity problem}

The auction in Bridge can be regarded as a CCP. Let $H_N$, $H_E$, $H_S$ and $H_W$ denote the hands of the players. There exists an optimal contract for $W$ and $E$ which is a function of all the hands $f(H_N,H_E,H_S,H_W)$. The goal of the partners is to call the optimal contract, in other words: find the value of $f$, with as little communication as possible. There are some additional constraints which make the problem more challenging: (i) the amount of communication which the players are allowed to exchange depends on the value of $f$ and on the strategy of their opponents, (ii) the function $f$ may be hard to compute even if all the hands are known, (iii) it is difficult to compare the quantum and the classical strategies, since in most of the cases it is hard to find the optimal classical strategies.

\begin{figure}[t]
\centerline{\includegraphics[width=0.9\columnwidth]{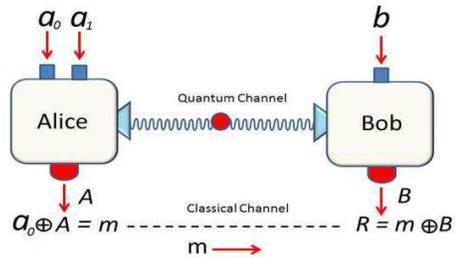}}
\caption{Quantum protocol of Bridge CCP. Alice (playing East) holds two bits of information $a_i$ with $i=0,1$ (defined by her hand). Bob (playing West) chooses between $b=0$ or $1$, which denotes in which bit of Alice he is interested.   Alice chooses her measurement setting to be $a=a_0\oplus a_1$. Bob sets his measurement according to his choice of $b$. After reading out her outcome $A$ (which is encoded as a bit value), Alice sends,  a one bit message to Bob, the value of  $m=A\oplus a_0$. Bob computes his guess  $R$ for the value of the bit he wants to know  by adding the message to his local measurement outcome $B$ (also encoded as a a bit value), that is $R=B\oplus m$.}
\label{scheme-Bridge}
\end{figure}

Nevertheless, one can show that the bridge scenario in figure  \ref{cards-Bridge} is equivalent to a CCP, and prove that the quantum strategies provide an advantage over classical ones. Let the bit $b$ represent  the type of information that Bob (playing West) is interested in (see Fig. \ref{scheme-Bridge}). $b=0$ corresponds to Bob being interested in the key cards, and $b=1$ to him being interested in $\sD$Q. For Alice (playing East) her bits will have the following meaning: the bit $a_0=0$ stands for 0 or 3 key cards, whereas $a_0=1$ means 1 or 4 of them; the bit $a_1=0$ means that Alice has $\sD$Q and $a_1=1$ means that she does not have one. The assignments of the particular values of the bits are arbitrary. For  simplicity we assume that for all the variables the value $0$ corresponds to the most probable situation. If Alice has $2$ or $5$ key cards (the bit $a_0$ is not defined) her answer is $5\sH$ or $5\spadesuit$ just like in the standard Blackwood. However, with the bits $a_0$ and $a_1$ well defined, it would be of an advantage for them to send a message to Bob allowing him to gain one of these values, or increase the probability of guessing them.  But,  she can send exactly one bit of information to Bob without exceeding the critical value of 5$\sD$. That is, her response can be 5$\sC$ or 5$\sD$.  Without knowing whether Bob is interested in $a_0$ or $a_1$ she can only, in the classical case, make a random choice, or they could agree beforehand that in such a case she sends, say $a_0$. This highly limits their strategies.

Clearly, their  task is a certain CCP. This can be put as follows:  Alice has a random string of two bits $a_0$ and $a_1$, while  Bob makes his independent choice, which of the two bits he wants to learn, that is he fixes $b=0,1$, to learn $a_b$ (see Fig. \ref{scheme-Bridge}).. But, Alice is allowed to send a single bit message $m$ to Bob ($m=0$ in the form of  5$\sC$ or $m=1$ as 5$\sD$), while he is not allowed to send any information.

For such CCPs one is usually interested in the worst case success probability, or the average probability (the success  is when Bob learns the correct value of $a_0$ or $a_1$). One usually assumes uniformly distributed inputs. However, the nature of Bridge is such that this assumption is not satisfied. Thus we must introduce the following figure of merit
\be \nonumber
I=\sum_{i,j,k}p(a_0=i,a_1=j,b=k)\times 
\\
P(R=a_b|a_0=i,a_1=j,b=k,m)
\ee
where $R$ is the value that Bob guesses for $a_b$, after receiving $m$, and $p(a_0,a_1,b)$ is the probability distribution of $a_0,a_1$ and $b$. In simple words, $I$ is the average  probability that Bob, after receiving the message $m$ correctly guesses the Alice's bit he wants to know.

To find the maximal value of $I$ achievable with classical resources, as in the case of any classical CCP,  it suffices to check all deterministic encodings of $a_0$ and $a_1$ into one bit message $m(a_0,a_1)$. The optimal deterministic strategy, that attains the maximum value of $I_c$ corresponds to  Alice in each run sending  $a_0$ (or $a_1$) as the message $m$. If Bob is interested in $a_0$ (alternatively, $a_1$), he gets its right value. However,  if he, in a given run, is  interested in $a_1$ (alternatively, $a_0$), his guess is the more probable value, which he infers from the known marginal distributions $p(a_0)$ and $p(a_1)$.  Thus one has
\begin{equation} \label{ic}
\begin{split}
I_C = \max \{  p(b=0) + p(b=1)&\max_i\{p(a_1=i|b=1)\}, \\
p(b=1)+& p(b=0)\max_i\{p(a_0=i|b=0)\} \}.
 \end{split}
\end{equation}

In the game of duplicate bridge to win does not mean to get more points than your opponents. It means to get more points than other pairs playing with the same cards. Because finding the optimal bid dramatically increases the amount of points awarded (see "Scoring and match points" section of the appendix), doing so is the necessary condition for winning. Therefore, the probability of guessing $a_b$ is the probability of winning and is referred to as such in the rest of the paper.

\section{IV. Quantum Bridge}

As we have explained in the previous section, the task that Alice and Bob have to perform in this situation reduces to communicating to the receiver one of the two bits. The only problem is that it is the receiver that chooses what he is interested in and the communication is restricted to a single bit. This kind of task is called a random access code and it is known that quantum information theory can deal with it more efficiently.

Let Alice and Bob make measurements on an entangled state. Each of them can choose one of two observables (Alice's choice is denoted by $a$ and Bob's by $b$) with binary outcomes $A$ and $B$ respectively. Consider a following protocol based on a $2\to 1$ entanglement assisted random access code discussed in Ref. \cite{M-EARAC}:

Alice receives two bits $a_0$ and $a_1$. She chooses her measurement setting  according to  the value of $a=a_0\oplus a_1$. The probability of $a$ to be $0$ is equal to $p=p(a_0=0)p(a_1=0)+(1-p(a_0=0))(1-p(a_1=0))$. Bob's setting is simply defined by his input bit $b$ which can be $0$ with probability $q$. After reading out her outcome $A$, Alice prepares the message $m=A\oplus a_0$ which she transmits to Bob. He then computes his guess value of the required bit, $R$, by adding the message to his outcome: $R=B\oplus m$. A simple calculation shows that $R=a_b$ as long as $A\oplus B=ab$.

Therefore, in our protocol Bob gets the correct value of $a_b$ with probability $I$ equal to
\be
I=\sum_{a,b} p(a,b) P(A\oplus B=ab|a,b),
\ee
which can be put equivalently as
\be
Q=\frac{1}{2}+\frac{1}{2}\sum_{a,b} p(a,b) (-1)^{ab}E(a,b),
\ee
where the correlation function $E(a,b)$ is given by $P(A=B|a,b)-P(A\neq B|a,b)$. The expression $Q=\sum_{a,b} p(a,b) (-1)^{ab}E(a,b)$ is the left hand side of a biased
CHSH inequality considered in \cite{LLP10}.  Its the maximal
quantum value is given by  \cite{LLP10}
\be
Q=\sqrt{2}\sqrt{q^2 + (1-q)^2}\sqrt{p^2 + (1-p)^2}.
\ee
Thus, the maximal quantum value of $I_Q$ is
\beq \label{iq}
I_Q=\frac{1}{2}\left(1+ \sqrt{2}\sqrt{q^2 + (1-q)^2}\sqrt{p^2 + (1-p)^2}\right).
\eeq

The exact probability distribution $p(a_0,a_1,b)$ is difficult to estimate since it depends on the overall strategies of the partners and their opponents. We have asked the bridge expert Tommy Gullberg for his estimates which are\cite{tommy}: $p(a_0=0|b=0)\approx 0.5, p(a_1=0|b=1)\approx 0.55$ which implies $p\approx 0.5$. It is also reasonable to assume $q\approx 0.75$.
With these estimates the classical strategy (\ref{ic}) gives success probability 0.8875 while the quantum one (\ref{iq}) reaches 0.8953.

As we have explained before, the same quantum protocol may be used in many different instances of bidding and trick-taking. However, the probability distribution $p(a_0,a_1,b)$ may differ from case to case. Therefore, in the experimental section of the paper we have compared the performance of quantum and classical strategies for a wide range of parameters. Usually, $p(a_0=0|b=0)\approx p(a_1=0|b=1)\approx p$ and we use this approximation in all our figures.

Let us stress that the game of Bridge as a whole is {\em not} a communication complexity
problem. It only shares some properties of it, in certain clearly defined phases of the game. For example its rules do
not put any constraints on the amount of communication between the
partners but only on the amount of {\em communication about their cards}. They
can, for example ask about the other players strategies, discuss issues
with the referee or simply ask the other partner to repeat his last bid
if they did not hear it clearly. This means that, in the quantum strategy
the players can make measurements on entangled pairs, and announce if
they detected a particle, until both of them do. This information has
nothing to do with their cards. Then they can carry on with the auction.
This allows them to exploit the advantage of quantum states without
having to worry about the efficiency of their detectors, which in
standard communication complexity problems plays a crucial role. We also like to point out that in  bridge offers no communication between partners that is not also available for the opposing team and hence eavesdropping on the opposing team's messages is pointless.

Apart form the main advantage of having a larger probability of playing the optimal contract, or better efficiency in the defense play, Alice and Bob also have a less obvious advantage. The rules of Bridge forbid the partners to use a secret strategy. Its detailed description should be made available to the opponents before the game.
Moreover, the player making the bid after Alice (the sender) and before Bob (the receiver) can, before announcing his/her bid, ask the receiver what information did Bob get and he must provide all the information that he has obtained. However, in the quantum case the receiver can answer truthfully:
{\it I did not get any information yet because I haven't measured my system yet and I cannot do it now because the choice of my measurement depends on your forthcoming bid. } Therefore, using the quantum strategy not only helps the partners to behave more optimally, but also makes the game harder for the opponents, as the overtly conveyed message carries no information, until it is added to the result of the receiver.  Furthermore, in the quantum case the choice of what the receiver learns is delayed until the very last moment before his bid, which gives him more knowledge about the opponents' cards. This allows him to learn information which is more relevant as he knows more about its context. An advantage of this type is difficult to quantify and delaying the choice of measurement requires high detection rates, therefore we do not dwell on this subject anymore here.

\begin{figure}[t]
\centerline{\includegraphics[width=0.82\columnwidth]{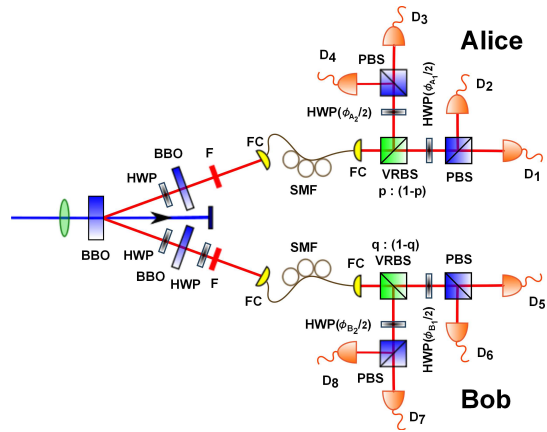}}
\caption{Experimental setup for Quantum Bridge CCP. UV light centered at wavelength of 390\, nm  are focused inside a 2\,mm thick BBO ($\beta$ barium borate) nonlinear crystal, to produce photon pairs. Half wave plates (HWP) and two 1\, mm thick BBO crystals are used for compensation of longitudinal and transverse walk-offs. The emitted photons are coupled into 2\, m single mode optical fibers (SMF) and passed through a narrow-bandwidth interference filters (F) ($\Delta\lambda=1$\,nm). Alice ( or the $E$ parter) uses $(p:1-p)$ variable-ratio beam splitter (VRBS) for her measurement basis choice with the probabilities $(p)$ and $(1-p)$ (corresponding to her input value $a$, see text) for the first and second base choice.  Her measurement observables $A_1$ and $A_2$ are realized by HWP oriented by $\phi_{A_1}$ and $\phi_{A_2}$ respectively. Bob uses $(q:1-q)$ variable ratio beam splitter (VRBS) for his basis choice with the probabilities $(q)$ and $(1-q)$ (corresponding to his input value $b$), for the first and second basis.  His observables $B_1$ and $B_2$ are realized by HWP oriented by $\phi_{B_1}$ and $\phi_{B_2}$ respectively. The polarization measurements for Alice and Bob were performed using polarizing beam splitters (PBS) and single photon detectors (D).}
\label{setup-CHSHgame}
\end{figure}
\section{V. Experimental realization}
Let's describe our experiment, which  demonstrates the quantum violation of the classical bound of (\ref{iq}) for various values of $p$ and $q$. At the same time it is the experimental realization of a biased nonlocal game \cite{LLP10,W10}, a quantum CCP, and the quantum Bridge. The optimal state for all these tasks is a two qubit maximally entangled state and the  measurements of the parties are the ones given in \cite{LLP10}:
\begin{eqnarray}
A_1&=&\frac{\sigma_x(q+(1-q)\cos\beta)+\sigma_z(1-q)\sin\beta }{\sqrt{(q+(1-q)\cos\beta)^2+(1-q)^2\sin^2\beta}},
\nonumber\\
A_2&=&\frac{\sigma_x(q-(1-q)\cos\beta)-\sigma_z(1-q)\sin\beta}{\sqrt{(q-(1-q)\cos\beta)^2+(1-q)^2\sin^2\beta}},
\nonumber\\
B_1&=&\sigma_x,\nonumber\\
B_2&=&\sigma_x\cos\beta  + \sigma_z\sin\beta,\nonumber\\
 \ket\psi&=&{1\over \sqrt 2} \left(\ket 0\ket 0 +\ket 1\ket
1\right),
\end{eqnarray}
where $\sigma_z$ and $\sigma_x$ are the standard Pauli operators, and
\begin{eqnarray}
\cos\beta=
\frac{1}{2}\frac{(q^2+(1-q)^2)(p^2-(1-p)^2)}{q(1-q)(p^2+(1-p)^2)}.
\end{eqnarray}
\begin{figure}[t]
\centerline{\includegraphics[width=0.82\columnwidth]{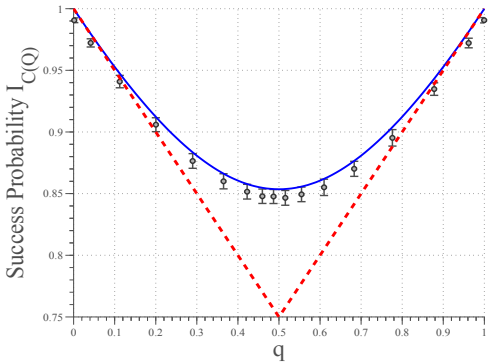}}
\caption{Experimental results for Quantum Bridge: (a) the classical $I_{C}(q)$ (dashed line), quantum $I_{Q}(q)$ (continuous line), and the experimental data points observed for the success probability $I$ for $p(a_0=0|b=0)= p(a_1=0|b=1)=p=0.5$. For $q=0.5$, the obtained quantum success probability is $0.848\pm 0.005$ (the corresponding classical value is $0.75$). For the value of $q=0.75$, which we estimate to correspond to a common situation in Bridge, the quantum and classical values are 0.895 (the experimental quantum value is $0.887 \pm 0.005$) and $0.875$ respectively.}
\label{results-p-CHSHgame}
\end{figure}

\begin{figure}[t]
\centerline{\includegraphics[width=0.82\columnwidth]{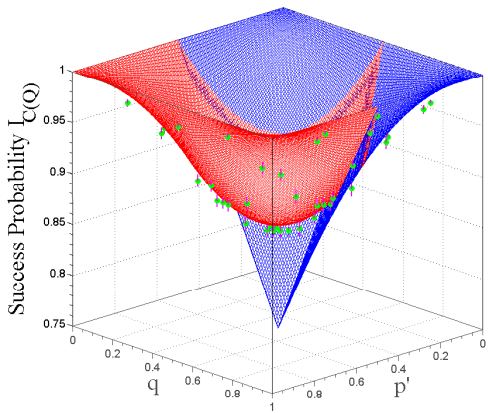}}
\caption{Experimental results for quantum CCP protocol for the whole spectrum of quantum strategies. The classical $I_{C}(p,q)$ (blue plot), quantum $I_{Q}(p,q)$ (red plot), and experimental data (green points)  for the success probability $I$ as functions of  $p$ and $q$. The classical value of $I_C(p,q)$ is calculated from (\ref{ic}) under the assumption that $p(a_0=0|b=0)= p(a_1=0|b=1)=p'$, which leads to $p=p'^2+(1-p')^2$. (Only half of $I_C(p,q)$ plot is shown to make the $I_Q(p,q)$ plot and the experimental data points visible).}
\label{results-pq-CHSHgame}
\end{figure}
We have realized these quantum protocols by using polarization entangled pairs of photons $|\phi^{+}\rangle = (\ket{HH} + \ket{VV})/\sqrt{2}$. In the experiment, UV light centered
at wavelength of 390\, nm  was focused inside a 2\,mm thick BBO ($\beta$ barium borate) nonlinear crystal. Photon pairs, due to a  degenerate emission of type-II spontaneous parametric down-conversion, are collected in two spatial modes $a$ and $b$. Half wave plates (HWP) and two 1\, mm thick BBO crystals are used for compensation of longitudinal and transversal walk-offs. The emitted photons were coupled into 2\, m single mode optical fibers (SMF) and passed through a narrow-bandwidth interference filters (F) ($\Delta\lambda=1$\,nm) to secure well defined spatial and spectral emission modes (see
Fig.~\ref{setup-CHSHgame})\cite{SPDC}.

Alice uses $(p:1-p)$ variable ratio beam splitter (VRBS) for her measurement basis choice with the probabilities $(p)$ and $(1-p)$ for the first and second basis choice. Her measurement observables $A_1$ (corresponding to $a= 0 $) and $A_2$ (corresponding to $a=1$) are realized by HWP oriented by $\phi_{A_1}$ and $\phi_{A_2}$ respectively. Bob uses $(q:1-q)$ variable ratio beam splitter (VRBS) for his basis choice with the probabilities $(q)$ and $(1-q)$ for the first and second basis choice. His measurement observables $B_1$ and $B_2$ (corresponding to $b=0$ and 1 respectively) are realized by HWP oriented by $\phi_{B_1}$ and $\phi_{B_2}$ respectively (see fig. \ref{setup-CHSHgame}) such as:
\begin{eqnarray}
\tan{\phi_{A_1}}&=& \frac{(1-q)\cos{(\pi/2 - \beta)}}{q+(1-q)\sin{(\pi/2 - \beta)}},
\nonumber\\
\tan{\phi_{A_2}}&=& \frac{-(1-q)\sin{(\pi/2 - \beta)}}{q-(1-q)\cos{(\pi/2 - \beta)}},
\nonumber\\
\phi_{B_1}&=& \pi/4,
\nonumber\\
\phi_{B_2}&=& \pi/2 - \beta.
\end{eqnarray}

The polarization measurement was performed using polarizing beam splitters (PBS) and single-photon detectors (D) placed at the two output modes of the PBS. Our detectors are  actively quenched Si-avalanche photodiodes.  All single detection events  were registered using a VHDL programmed multichannel coincidence logic unit, with a time coincidence window of 1.7 ns. The measurement time for each setting was $200$ seconds.

We have tested the  protocols for different probabilities $p$ and $q$ by changing  the transmission coefficients of Alice's and Bob's VRBS. Fig.
\ref{results-p-CHSHgame} shows the classical $I_{C}(q)$ (dashed line), quantum $I_{Q}(q)$ (continuous line), and experimental data points observed for the success probability versus $q$ for the value of $p=0.5$. Fig. \ref{results-pq-CHSHgame} shows $3D$ plot for the classical
$I_{CS}(p,q)$, quantum $I_{QS}(p,q)$, and experimental data  observed for the success probability of CCP protocols versus $p$ and $q$. Our results
are in very good agreement with the theoretical predictions and clearly demonstrate the advantage of the quantum strategy over the classical ones.


\section{VI. Conclusion}

In summary, we report an experimental realization of a quantum Bridge CCP protocol, which is the first demonstration of  a quantum CCP usable in a real-life scenario. This was possible because we show that our quantum Bridge protocol corresponds to a biased nonlocal CHSH game,  which in turn is equivalent to a $2\to 1$ quantum random access code. Thus, our experiment is also a realization of an entanglement assisted random access code.  We also establish links between game theory, communication complexity, and quantum physics. Such quantum games and communication complexity protocols and their links can be generalized to $n$ parties scenario.  Concerning  Bridge, it is up to the World Bridge Federation to decide whether to allow quantum resources and encoding strategies in championships. A positive decision would  make this technique the first commonplace application of quantum communication complexity. A negative one would forbid quantum strategies  and thus, would constitute the first  ever regulation of quantum resources sport. The results reported here will contribute to deeper understanding of the possible impact of quantum resources on information and communication technologies.

\section{acknowledgments}
The authors thank Johan Ahrens and Tommy Gullberg for discussions. This work was supported by the Swedish Research Council (VR),the Linnaeus Center of Excellence ADOPT, TEAM programme of foundation for Polish Science (FNP), NCN grant 2013/08/M/ST2/00626, QUASAR (ERA-NET CHIST-ERA 7FP UE) and UK EPSRC.

\section{Appendix: The Rules of Duplicate Bridge}

Duplicate bridge is a metagame. The same bridge deal (i.e. the specific arrangement of the 52 cards into the four hands) is played at each table and scoring is based on relative performance. This reduces the element of chance while heightening the one of skill. Now we give the brief description of the game. More details can be found in e.g \cite{Bridge}.
\subsection{Introduction}
Duplicate bridge is one of the most popular card games, recognized by the International Olympic committee as a sport (only two "mind sports": bridge and chess are recognized) and governed by the World Bridge Federation for international competitions. It uses a standard deck of 52 cards and is played by exactly four players. The players are usually named according to their seats directions as East, West, North and South. Among these West and East form a partnership or a pair competing against North and South.

A tournament consists of several games. In each game the cards are distributed between the players. In rubber bridge a more casual version of the game this distribution, called a deal, heavily influences the winning probability. However, in duplicate bridge previously prepared deals are stored in bridge boards - simple four-way card holders. They are used to enable each player's hand to be passed intact to the next table that must play the same deal, and final scores are calculated by comparing each pair's result with others who played the same hand.

Each game consists of two main phases:
\begin{enumerate}
	\item The auction, also called bidding for a contract;
	\item Playing the hand, when the pair that won the auction tries to take enough tricks to make the contract.
\end{enumerate}

\subsection{Bidding in Bridge}

In order to understand how to bid, first one should know the ranking of cards and suits. The deck of 52 cards consists of 4 suits: spades $\sS$, hearts$\sH$, diamonds $\sD$ and club $\sC$. Spades $\sS$ are ranked as the highest and the next suit is hearts $\sH$, together they are called the major suits. They are followed by minor suits with diamonds $\sD$ ranked higher than clubs $\sC$. This ranking is important in biding and scoring at the end of the game. In a suit, cards are ranked from 2 being the lowest to ace being the highest.

The bidding phase starts with the dealer (one player is marked as dealer on each bridge board) and rotates around the table clockwise with each player making a call. A call is limited to a vocabulary of 38 words or phrases consisting of:
\begin{itemize}
\item a Bid which states a level and a denomination; a denomination can be any of the four suits or NT (which stands for No Trump); given 7 levels of bidding and 5 denominations, there are 35 possible bids;
\item Double, only available when the last bid was made by an opponent;
\item Redouble, only available after opponent's double;
\item Pass, when unwilling or unable to make one of the three preceding calls.
\end{itemize}
This vocabulary is further limited by the requirement that each bid has to be sufficient, i.e. it has to have either a higher level than a previous bid or the same level and a higher denomination. NT is higher than any of the suits. For example, if the last bid was 3$\sS$ the next one can be 3NT or 4$\sC$ but not 3 $\sC$.

If three players in a row pass, the auction is over. A pair is said to have won the auction if the last bid was made by one of its players. This partnership is called the declaring side.  The player on the declaring side who, during the auction, first stated the denomination of the final bid becomes "declarer," the declarer's partner becomes the "dummy," and the opposing side become the "defenders." The final bid also becomes the "contract". The declaring side promises to take, during the next phase, the number of tricks greater or equal to the level of the contract plus 6 and the denomination becomes the trump (in case of denomination NT, there is no trump).

\subsection{Playing the hand}

To begin play, the defender on the declarer's left makes the opening lead by placing his selection face up on the table. The dummy then spreads his hand on the table so that it is visible to every other player. The players play clockwise around the table placing their cards on the table, and each must "follow suit" (that is, play a card of the suit lead to the trick) if able. A player that cannot follow suit may either "ruff" (play a trump) if there is a trump suit or "sluff" (play a card of any other suit). The player that plays either the highest trump or, in a trick that contains no trumps, the highest card of the suit led to the trick (1) wins the trick for its side and (2) proceeds to lead to the next trick. The declarer directs the play of cards from the dummy in addition to playing cards from his own hand. The play continues until all thirteen tricks are played. Then the score is calculated.

\subsection{Scoring and match points}

If the declaring side won the number of tricks specified by contract (or more) they are awarded some points and their opponents get exactly the same amount of negative ones. If they won less the situation is reversed.

If the declaring side made their contract, they get the points for the following:
\begin{itemize}
\item Odd tricks - the number of tricks specified by the level of the contract. The amount of points depends on the level of the contract, denomination and whether it was doubled or redoubled. The points awarded for this are called contract points.
\item Overtricks - if the partnership won more tricks than contract specified they are counted as overtricks. They are worth less points than odd tricks. The amount of points here also depends on denomination and whether the contract was doubled or redoubled but also on vulnerability. Whether or not the partnership is vulnerable is marked on the bridge board.
\item Slam bonus - if the partnership bids and wins 12 or 13 tricks they are awarded a huge bonus which depends on their vulnerability.
\item Double bonus - A bonus for making the contract that has been doubled or redoubled.
\item Game bonus - A small bonus for making the contract is awarded. A much larger one is given if the partnership scored more than 100 contract points. Even larger if they were vulnerable.
\end{itemize}

If the declaring side failed to make their contract, their opponents get points for each undertrick. The amount of points is larger than for overtricks or even odd tricks and depends on vulnerability and whether the contract was doubled or redoubled.

The tables with the exact point values can be found e.g. here \cite{scoring}. An important implication of these scoring rules is that finding the optimal contract is crucial for the game. This stems from the fact that overtricks are worth less than odd tricks, game bonus is awarded only for contract points and slam bonus only applied if the parties bid for a slam. For example, let us consider a case where the declaring side won 9 tricks with no trump while being vulnerable. If their contract was 2NT they are awarded 70 contract points plus 30 for one overtrick plus 50 of game bonus, which gives them 150 total. If their contract was 3NT they get 100 contract points and a gem bonus of 500, which gives them 600 in total. If their contract was 4NT they had one undertrick and they get 100 penalty points. It is therefore crucial for the declaring side neither to under- or overestimate their ability to win tricks.

However, these points do not influence the final position of the partnership that scored them directly. Instead all the pairs that played the same deal are compared (the pairs that sat at North-South and East-West positions are compared separately). Then 2 match points (MPs) are awarded for each partnership that scored less points with the same deal. $\frac{1}{2}$ MP is awarded for each partnership that scored the same amount. Therefore, scoring 90 MPs when everyone else made only 70 is as good as scoring 2000.

\subsection{Bidding strategies}

In duplicate bridge auction phase is much more important than playing the hand. Often, the second phase is reduced to the declarer revealing his hand and claiming that he will take exactly the amount of tricks specified in the contract and the defenders agree ending the game.

The auction phase is all about communication. The purpose of some early bids may be to exchange information rather than to set the final contract. For most players, many calls (bids, doubles and redoubles, and sometimes even passes) are not made with the intention that they become the final contract, but to describe the strength and distribution of the player's hand, so that the partnership can reach an informed conclusion on their best contract, and/or to obstruct the opponents' bidding. The set of agreements used by a partnership about the meaning of each call is referred to as a bidding system, full details of which must be made available to the opponents; 'secret' systems are not allowed. An opponent can ask the bidder's partner to explain the meaning of the call.

There are around $5.36\times 10^{28}$ different deals possible which makes bidding strategies very complex, especially that they have to take into the account possibility of the opponents interfering. To make it possible for humans to play the game, they are divided into conventions - sub-strategies that are applicable only in certain situations, e.g. Cappelletti convention is a strategy for interrupting opponents communication while sending the partner information about the strongest suit; it can be used only after bid 1NT by the opponents by a player with moderately strong cards.

Because making a contract for more than 100 contract points or a slam leads to large amounts of bonus points there are multiple conventions designed to check if such contracts are possible to make. One of the most popular is Roman Key Blackwood described in the main text. All the conventions, apart for the ones for interrupting the opponents, are designed to convey as much information as possible while keeping the final bid as low as possible. This enables the players to stay at safe level in case they find out that their cards are not that good or to squeeze in additional rounds of communication before concluding the auction. Therefore they bear resemblance to communication complexity problems.

\subsection{Strategies for playing the hand}

The strategies for playing the hand are very different for the declaring side and the defenders. This stems from the fact that dummy's cards are visible to everyone and they are being played by the declarer. This leads to asymmetry. The declaring side is reduced to a single player and half of its cards are known to the opponents. The defenders, on the other hand, face a problem of coordinating their actions. Usually, they do not have many opportunities to exchange information during the auction so they have to use the cards they play to send signals. One of the most common conventions is called Smith Peter. When one of the defenders plays the opening lead the other plays a low spot card (2-5) if he would like his partner to play this suit once more and high spot card (6-10) if he wants to  discourage his partner from playing this suit. This strategy is called signalling.

Again, the communication is limited because the players aim at establishing a joint defence strategy as soon as possible and their freedom of choosing the signals is constrained by the cards they have. Therefore, the quantum protocol described in the main text can be adapted also to defence play. An example of this is presented in the next part of the appendix.

Duplicate bridge is one of the most complex games and this short description presents only the very basics of it. Interested reader should read any of the numerous books available on this subjects and try to play the game himself/herself.

\section{Appendix: Quantum strategies for defence play}
\begin{figure}[t]
\centerline{\includegraphics[width=0.60\columnwidth]{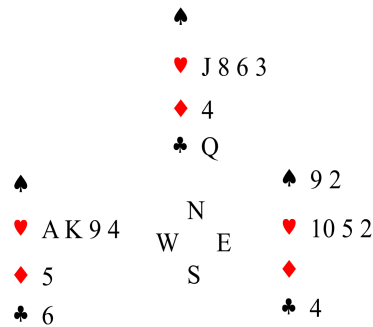}}
\caption{Cards for defense play and signaling.}
\label{cards-Bridge_defence}
\end{figure}
The quantum protocol described in the main text can also be used for defense play and signaling. To see this consider the cards given in figure \ref{cards-Bridge_defence}.
In this particular game of Bridge, the W-E team is the defending team and N is the dummy.
In this given scenario, let assume that W discards the $\sH A$ against S's $4\sD$ contract. After N's response, East plays the $\sH 10$ to signal a discouraging attitude towards the played suit (see Fig \ref{cards-Bridge_defence}). For W, it is interesting to know whether E possesses the $\sH Q$  or not. From the viewpoint of W, E can have $\sH 10$ as the highest card in the suit or East might have e.g. $\sH Q$ and still play $\sH 10$ to signal a negative attitude. If East has the queen, West should discard the $\sH 9$ so that East can play the $\sH Q$. If East does not have the relevant card, there is a good risk of letting the declarer win the 10 tricks needed to fulfill the contract. In conclusion; West needs information about the $\sH Q$.
In order to extract the information about the queen, we once again use the given quantum protocol with the following variable definitions. If $a_0=0$, encourages suit. If $a_0=1$, discourages suit. If $a_1=0$, odd number of cards. If $a_1=1$, even number of cards
With $q$ defined as the probability of W being interested in the attitude towards the suit we may define the message $m$ as: if $m=0$, then play a small card, if $m=1$, then play a high card. Thus, the quantum protocol has once again improved the conditions of playing the best card.
Generally, whenever communication between the partners is restricted to one single bit and there exists a non-zero probability that the partner might be interested in values of one of {\em two} bits of information, the protocol gives a quantum advantage.

\end{document}